\pdfoutput=1

\documentclass[preprint,authoryear,5p]{elsarticle}

\usepackage{graphicx,amssymb}
\usepackage{epsf,psfig}
\usepackage{natbib}
\usepackage[unicode]{hyperref}

\journal{Mechanics Research Communications}

\begin{document}

\begin{frontmatter}
\title{Order and chaos in a three dimensional galaxy model}

\author[cj]{Christof Jung}
\ead{jung@fis.unam.mx}

\author[eez]{Euaggelos E. Zotos\corref{cor1}}
\ead{evzotos@physics.auth.gr}

\cortext[cor1]{Corresponding author}

\address[cj]{Instituto de Ciencias F\'{i}sicas, \\
Universidad Nacional Aut\'{o}noma de M\'{e}xico \\
Av. Universidad s/n, 62251 Cuernavaca, Mexico}

\address[eez]{Department of Physics, School of Science, \\
Aristotle University of Thessaloniki, \\
GR-541 24, Thessaloniki, Greece \\}

\begin{abstract}
We explore the orbital dynamics of a realistic three dimensional model describing the properties of a disk galaxy with a spherically symmetric central dense nucleus and a triaxial dark matter halo component. Regions of phase space with regular and chaotic motion are identified depending on the parameter values for triaxiality of the dark matter halo and for breaking the rotational symmetry. The four dimensional Poincar\'{e} map of the three degrees of freedom system is analysed by a study of its restriction to various two dimensional invariant subsets of its domain.
\end{abstract}


\end{frontmatter}

\section{Introduction}
\label{intro}

To present a global panorama of the dynamics of a system by plotting Poincar\'{e} maps has a long tradition for two degrees of freedom (2-dof) systems where these maps live on a 2 dimensional domain (for a good explanation of the idea of Poincar\'{e} maps see chapter 6 in \citet{J90}). Unfortunately things are not so easy for more degrees of freedom. Already for 3-dof the domain of the Poincar\'{e} map has dimension 4 and a graphical representation of the map is impossible. However, such 4 dimensional maps usually have the property that there are 2 dimensional invariant surfaces in the domain, i.e. surfaces with the property that initial conditions on these lower dimensional surfaces lead to trajectories which lie completely in this lower dimensional surface. And then it is easy to present graphically the restriction of the map to any one of such surfaces. Such restricted 2 dimensional maps are a great help to understand important properties of the full 4 dimensional map.

To be useful such 2 dimensional invariant surfaces must have some kind of robustness property, i.e. they must survive parameter changes such that we can follow the perturbation scenario in the restricted map. There are two types of surfaces with this property. First, the so called normally hyperbolic invariant manifolds (abbreviated NHIMs, for their general properties see \citet{W94}), however, it seems that these objects do not play any important role in our present galaxy model. Second, lower dimensional surfaces may be invariant because of symmetry reasons, and in the present galaxy model there are various surfaces of this type which will be very useful in the following.

A further contribution to the understanding of 4 dimensional maps comes from a possible partially integrable limit case of the full system and from the corresponding possibility to build up the 4 dimensional map as stack of 2 dimensional reduced maps. Because of its importance for the present work let us give a short summary of the stack construction in a form appropriate for the case found in the dynamics of the galactic potential. Assume a Hamiltonian 3-dof system where the unperturbed case has a rotational symmetry around the $z$-axis and where correspondingly $L$, the $z$-component of the angular momentum, is conserved. Then the unperturbed system can be reduced to a 2-dof system which depends parametrically on the value of $L$. This leads to a 1 parameter family of reduced Poincar\'{e} maps each one acting on a 2 dimensional domain. From the reduced maps we obtain the non-reduced Poincar\'{e} map acting on its 4 dimensional domain in a two step process. First, we pile up the continuum of 2 dimensional reduced maps to a 3 dimensional stack where $L$ acts as stack parameter. Second, we form the Cartesian product of this pile with a circle representing the cyclic angle canonically conjugate to the conserved quantity $L$. The result is a 4 dimensional construction. The full 4 dimensional Poincar\'{e} map (still for the unperturbed case) acts on the resulting 4 dimensional domain as reduced map in each invariant horizontal leaf of constant $L$ and in addition applies a rotation to each copy of the circle. If we now add a perturbation which destroys the rotational symmetry around the $z$ axis then the invariant foliation into the horizontal leaves is destroyed. However, if the
perturbation conserves some discrete symmetry then the 4 dimensional map will have corresponding invariant 2 dimensional surfaces $S$ also after the perturbation. These 2 dimensional surfaces do not coincide with any one of the invariant 2 dimensional horizontal leaves of the unperturbed stack, they are transverse to the stack structure. Most important, the dynamics restricted to $S$ is sensitive to the perturbation and can be exploited to obtain an understanding of the perturbation scenario for the full 3-dof system. The restricted map on $S$ visualizes the decay of the stack structure under the perturbation. The idea works the same if the role of $S$ is taken over by a NHIM. For all details of this approach and for examples of the stack construction see \citet{JMSZ10,GJ12,GDJ14}. For the analysis of a 3-dof molecular system taking advantage of this procedure see \citet{LRJ15}. The present article shows that the method is equally successful for the investigation of the dynamics of a 3-dof galaxy model.

In order to explore the orbital dynamics of galaxies one should build first suitable models describing sufficiently and realistically the properties of the galaxy. Usually observations provide the necessary information on the construction of the dynamical models. A galactic model can be characterized as successful and realistic only if the derived results agree with the corresponding observational data. In most of the cases, the galaxy models are either spherical or axially symmetric. For instance, in a spherically symmetric potential all three components of the angular momentum and of course the total angular momentum are conserved. Therefore, the motion of the stars is plane and takes place in the plane perpendicular to the vector of the total angular momentum. Spherically symmetric models for galaxies were studied by \citet{D93,RDZ05,Z96}. In an axially symmetric potential on the other hand, only the $z$ component of the total angular momentum is conserved. Many previous papers are devoted to the distinction between order and chaos in axially symmetric potentials (see e.g., \citet{Z11,Z12,ZC14}). Furthermore, axially symmetric galaxy models were presented and examined in \citet{CdZ99}.

Another interesting category of galactic potentials are the so-called composite galaxy models. In those models the potential is multi-component and each component describes a different part of the system. Composite axially symmetric galaxy models describing the motion of stars in our Galaxy were also studied by \citet{B12}. In these models the gravitational potential is composed by three superposed disks: one representing the gas layer, one the thin disk and one for the thick disk. Recently in \citet{TS07}, a class of realistic triaxial models for galaxies was provided. In particular, the authors extended an earlier method proposed by \citet{TG05} to three-dimensional systems by replacing the radial with an ellipsoidal symmetry in the total mass density. Moreover, triaxial galaxy models were also constructed by \citet{BSB07} and \citet{MKDS04}.

The layout of the paper is as follows: Section \ref{galmod}, contains a detailed presentation of the structure and the properties of our galactic gravitational model. In Section \ref{numres}, we construct Poincar\'{e} maps in order to investigate the orbital properties of the 6 dimensional phase space. The paper ends with Section \ref{disc}, where the discussion and the main conclusions of our numerical analysis are presented.

\section{Presentation of the galactic model}
\label{galmod}

The total gravitational potential $\Phi(x,y,z)$ is three-dimensional and it consists of three components: a central, spherical component $\Phi_{\rm n}$, a flat disk $\Phi_{\rm d}$ and a triaxial dark matter halo potential $\Phi_{\rm h}$.

The spherically symmetric nucleus is modeled by a Plummer potential \citep{BT08}
\begin{equation}
\Phi_{\rm n}(x,y,z) = \frac{-G M_{\rm n}}{\sqrt{x^2 + y^2 + z^ 2 + c_{\rm n}^2}}.
\label{Vn}
\end{equation}
Here $G$ is the gravitational constant, while $M_{\rm n}$ and $c_{\rm n}$ are the mass and the scale length of the nucleus, respectively. Here we must point out, that potential (\ref{Vn}) is not intended to represent a black hole nor any other compact object, but a dense and massive bulge. Therefore, we don't include any relativistic effects.

In order to model the disk we use the Miyamoto-Nagai potential \citep{MN75}
\begin{equation}
\Phi_{\rm d}(x,y,z) = \frac{-G M_{\rm d}}{\sqrt{x^2 + y^2 + \left(s + \sqrt{h^2 + z^ 2}\right)^2}},
\label{Vd}
\end{equation}
where $M_{\rm d}$ is the mass of the disk, while $s$ and $h$ are the horizontal and vertical scale lengths of the disk.

\begin{figure}[!tH]
\begin{center}
\includegraphics[width=\hsize]{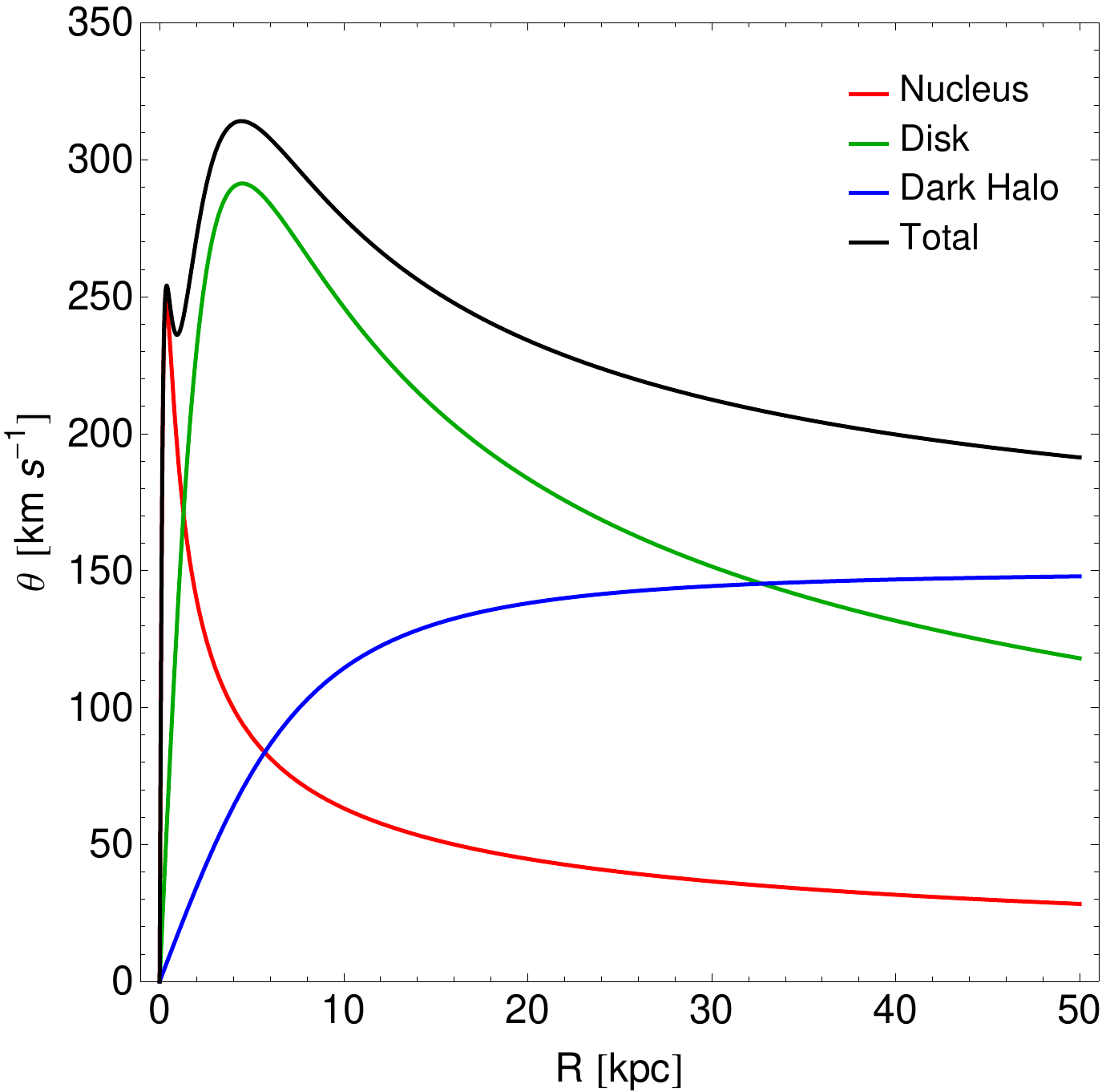}
\end{center}
\caption{A plot of the rotation curve in our galactic model. We distinguish the total circular velocity (black) and also the contributions from the central massive nucleus (red), the disk (green) and that of the dark matter halo (blue).}
\label{rotvel}
\end{figure}

For the description of the properties of the dark matter halo we use the logarithmic potential
\begin{equation}
\Phi_{\rm h}(x,y,z) = \frac{\upsilon_0^2}{2} \ln \left(x^2 + \alpha y^2 + \beta z^2 + c_{\rm h}^2 \right),
\label{Vh}
\end{equation}
where $\alpha$ and $\beta$ are the flattening parameters along the $y$ and $z$ axes, respectively, $c_{\rm h}$ is the scale length of the halo, while the parameter $\upsilon_0$ is used for the consistency of the galactic units. The choice for the logarithmic potential was motivated for several reasons: (i) it can model a wide variety of shapes of galactic haloes by suitably choosing the parameter $\beta$. In particular, when $0.1 \leq \beta < 1$ the dark matter halo is prolate, when $\beta = 1$ is spherical, while when $1 < \beta < 2$ is oblate; (ii) it is appropriate for the description of motion in a dark matter halo as it produces a flat rotation curve at large radii (see Fig. \ref{rotvel}); (iii) it allows for the investigation of flattened configurations of the galactic halo at low computational costs; (iv) the relatively small number of input parameters of Eq. (\ref{Vh}) is an advantage concerning the performance and speed of the numerical model; and (v) the flattened logarithmic potential was utilized successfully in previous works to model a dark matter halo component (e.g., \citet{H04a,RPT07,Z14}).

We use a system of galactic units, where the unit of length is 1 kpc, the unit of mass is $2.325 \times 10^7 {\rm M}_\odot$ (solar masses) and the unit of time is $0.9778 \times 10^8$ yr (about 100 Myr). The velocity units is 10 km/s, the unit of angular momentum (per unit mass) is 10 km kpc s$^{-1}$, while $G$ is equal to unity. The energy unit (per unit mass) is 100 km$^2$s$^{-2}$. In these units, the values of the involved parameters are: $M_{\rm n} = 400$ $c_{\rm n} = 0.25$, $M_{\rm d} = 7000$, $s = 3$, $h = 0.175$, $\upsilon_0 = 15$ and $c_{\rm h} = 8.5$, while $\alpha$ and $\beta$ are treated as variable parameters in the intervals $1 \leq \alpha \leq 2$ and $0.1 \leq \beta \leq 1.9$. The values of the disk and the nucleus were chosen with a Milky Way-type galaxy in mind (e.g., \citet{AS91}) and they secure positive density everywhere. Here we must point out that our total gravitational potential is truncated at $R_{max} = 50$ kpc, otherwise the total mass of the galaxy modeled by this potential is infinite which of course lacks of physical meaning.

For $\alpha = 1$ the total gravitational potential $\Phi(x,y,z)$ becomes axially symmetric $V(R,z)$, where $R^2 = x^2 + y^2$ and the $z$-component of the angular momentum is conserved. One of the most important quantities in disk galaxies is the circular velocity in the galactic plane which is defined as,
\begin{equation}
\theta(R) = \sqrt{R\left|\frac{\partial V(R,z)}{\partial R}\right|_{z = 0}},
\label{rcur}
\end{equation}
A plot of $\theta(R)$ for our galactic model\footnote{On the galactic plane applies $z=0$, so the rotation curve is the same regardless of the particular value of $\beta$.} is presented in Fig. \ref{rotvel} as a black curve. Moreover, in the same plot, the red line shows the contribution from the spherical nucleus, the green curve is the contribution from the disk, while the blue line corresponds to the contribution form the dark matter halo. It is seen that each contribution prevails in different distances form the galactic center. In particular, at small distances when $R\leq 2$ kpc, the contribution from the spherical nucleus dominates, while at mediocre distances, $2 < R < 32$ kpc, the disk contribution is the dominant factor. On the other hand, at large enough galactocentric distances, $R > 32$ kpc, we see that the contribution from the dark halo prevails, thus forcing the rotation curve to remain flat with increasing distance from the center. We also observe the characteristic local minimum of the rotation curve due to the massive nucleus, which appears when fitting observed data to a galactic model (e.g., \citet{GHBL10,IWTS13}).

The equations of motion for a test particle with a unit mass $(m = 1)$ are
\begin{equation}
\ddot{x} = - \frac{\partial \Phi}{\partial x}, \ \ \
\ddot{y} = - \frac{\partial \Phi}{\partial y}, \ \ \
\ddot{z} = - \frac{\partial \Phi}{\partial z},
\label{eqmot}
\end{equation}
where, as usual, the dot indicates derivative with respect to the time.

The Hamiltonian for the total potential $\Phi(x,y,z)$ reads
\begin{equation}
H = \frac{1}{2}\left(p_x^2 + p_y^2 + p_z^2 \right) + \Phi(x,y,z) = E,
\label{ham}
\end{equation}
where $p_x$, $p_y$ and $p_z$ are the momenta per unit mass conjugate to $x$, $y$ and $z$, respectively, while $E$ is the numerical value of the Hamiltonian, which is conserved.

\section{Numerical results}
\label{numres}

It is rather obvious what is the appropriate partially integrable limiting case of our galaxy model. It is the rotationally symmetric case with $\alpha = 1$. Then in cylindrical coordinates $(R, \phi, z)$ the potential is independent of $\phi$ and correspondingly the canonically conjugate action $L$, the $z$ component of the total angular momentum, is conserved. And we have a foliation of the system into leaves of constant $L$. First we treat the case of $E = 160$ and $\beta = 0.1$ in all detail and we give short remarks on other values of these two quantities later. Here, we have to point out that the energy level controls the size of the maps and particularly $x_{\rm max}$ which is the maximum possible value of the $x$ coordinate. We chose that energy level which yields $x_{\rm max} \simeq 15$ kpc.

From the functional form of the Hamiltonian in Eq. (\ref{ham}) it is also easy to see that the surfaces $S_x$ given by $x = p_x = 0$, $S_y$ given by $y = p_y = 0$, and $S_z$ given by $z = p_z = 0$ are invariant. Any initial condition starting in one of these surfaces leads to a trajectory which lies entirely in the respective surface.

\begin{figure*}[!tH]
\centering
\resizebox{0.8\hsize}{!}{\includegraphics{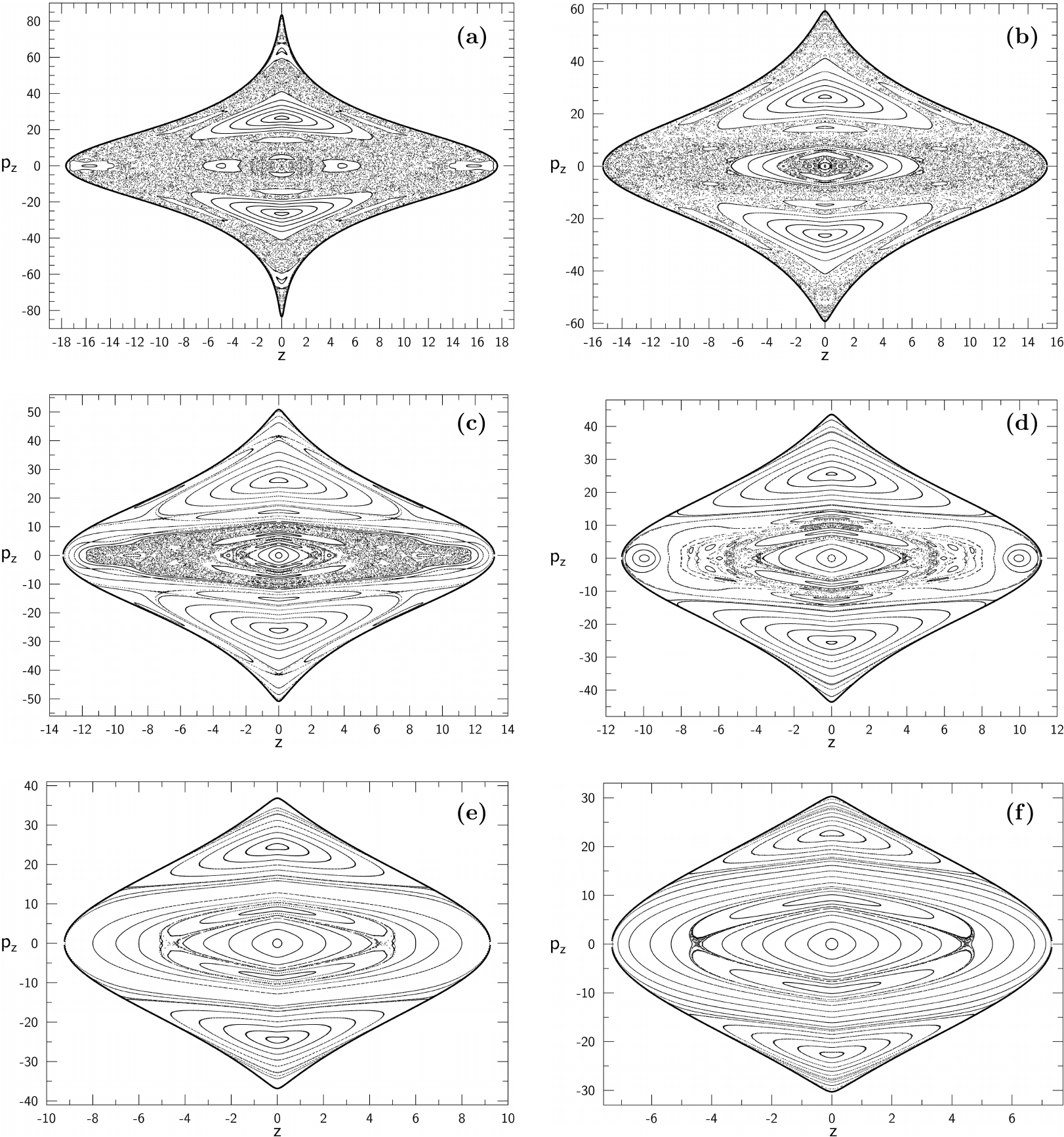}}
\caption{Plot of the Poincar\'{e} map $M_{\rm red}$ for the symmetric case with $\alpha = 1$ and $\beta = 0.1$ when $L$ varies. (a) $L = 0$; (b) $L = 30$; (c) $L = 60$; (d) $L = 90$; (e) $L = 120$; (f) $L = 150$.}
\label{map1}
\end{figure*}

Now we come to the question which intersection condition to use for the Poincar\'{e} map. It must be an intersection condition which is appropriate to investigate the restricted dynamics in the surfaces $S_x$, $S_y$ and $S_z$ and it must also be appropriate for the stack construction in the symmetric case. The best intersection condition we found fulfilling all these side conditions is the condition that the cylindrical radius $R$ is maximal, or equivalently that its conjugate momentum $p_R$ changes from positive to negative. The 4 dimensional Poincar\'{e} map generated with this intersection condition will be called $M$ in the following. The restrictions of this map to the invariant surfaces $S_x$, $S_y$ and $S_z$ will be called $M_x$, $M_y$ and $M_z$, respectively. As in most cases for realistic systems the transversality of the intersection condition is violated for a few important trajectories. In our case these trajectories include two simple ones. First the trajectory $\gamma_1$ which oscillates up and down along the z-axis with $R$ constantly zero. And second,
in the symmetric case the trajectory $\gamma_2$ which encircles the origin in a circle with constant value of $R$ and in the plane $z = 0$. As we will see later these periodic trajectories appear in some restricted maps as energetic boundaries.

\begin{figure}[!tH]
\begin{center}
\includegraphics[width=\hsize]{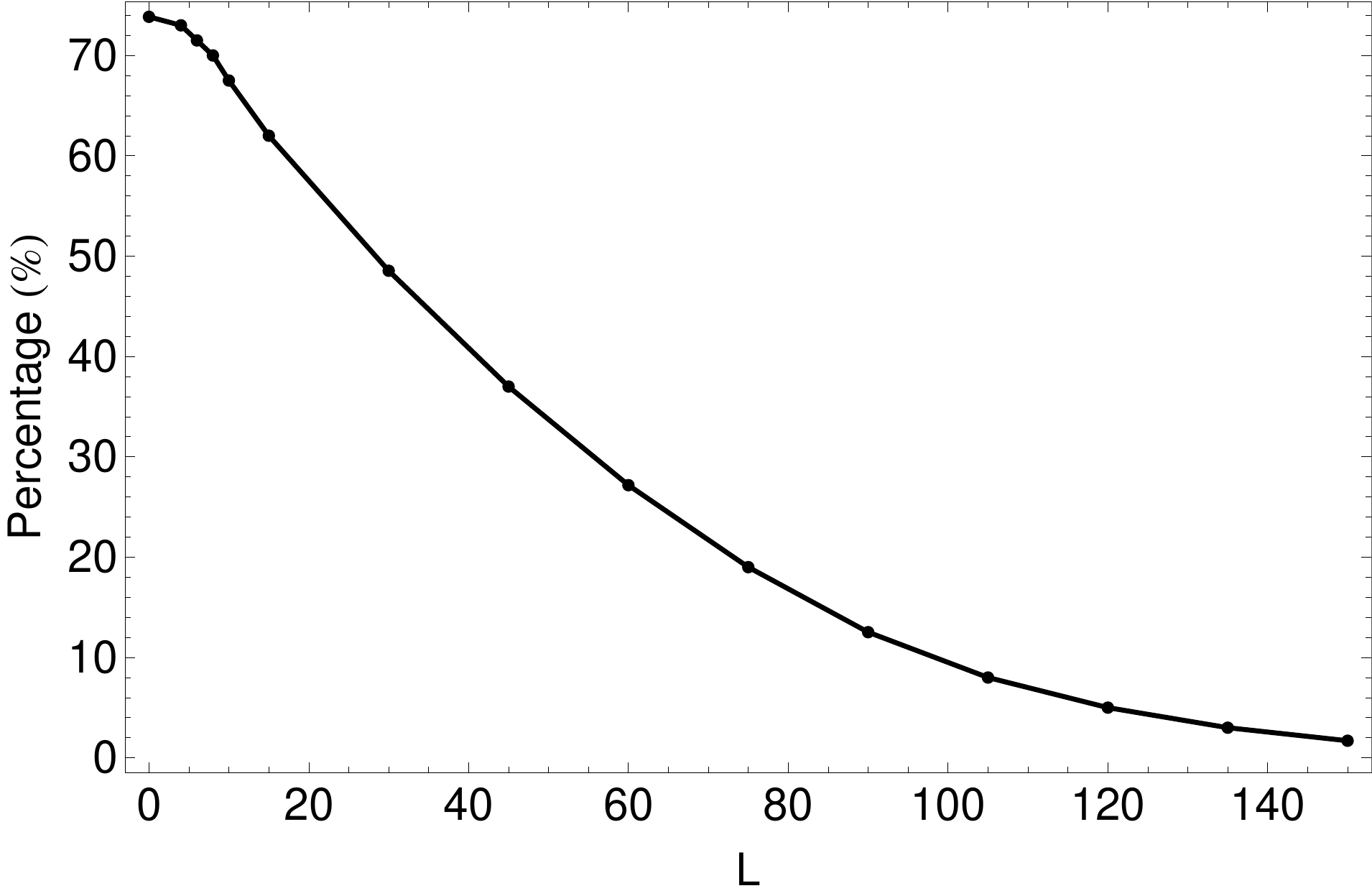}
\end{center}
\caption{Evolution of the relative fraction of area covered by chaotic trajectories as function of $L$ when $\alpha = 1$ and $\beta = 0.1$.}
\label{chaos}
\end{figure}

Next we explain the stack idea in detail. Consider the symmetric case with $\alpha = 1$, where in cylindrical coordinates the coordinate $\phi$ does not appear in the Hamiltonian and where $L$ is conserved and correspondingly can be treated as if it
would be a parameter. Because of the rotational symmetry the angle $\phi$ will be ignored for the moment. Then we have a foliation of the phase space and also of the domain of the Poincar\'{e} map into invariant leaves belonging to the various values of $L$ and each leaf can be treated as independent dynamical system. The reduced system is 2-dof ($R$ and $z$ degrees of freedom) and its reduced Poincar\'{e} map (called $M_{\rm red}$ in the following) acts on the 2 dimensional domain with the coordinates $z$ and $p_z$. In Fig. \ref{map1} we show the numerical plots of $M_{\rm red}$ for the examples of the $L$ values 0, 30, 60, 90, 120, and 150 in parts (a), (b), (c), (d), (e), and (f), respectively. We observe that with increasing
value of $L$ the domain of the reduced map shrinks and the map becomes more regular. For the value 160 of the total energy $E$ the maximal possible value of $L$ is found as $L_{max} \approx 245$ and when $L$ converges to this value from below then the domain of the reduced map shrinks to a single point, the origin. And in this limit case this single point represents the periodic trajectory $\gamma_2$. Fig. \ref{chaos} shows the relative fraction of area covered by chaotic trajectories as function of $L$. It demonstrates in a quantitative form how this fraction converges to 0 for increasing value of $L$. For the determination of the chaotic percentage we used an image processing numerical method which is explained in the Appendix. Of course, there are also corresponding leaves with negative values of $L$ and the opposite orientation of rotation around the $z$-axis. However, because the Hamiltonian is even in $L$ these cases are nothing new and only are mirror images of the cases with positive values of $L$. Therefore, there is no need to show numerical examples for negative values of $L$.

\begin{figure}[!tH]
\begin{center}
\includegraphics[width=\hsize]{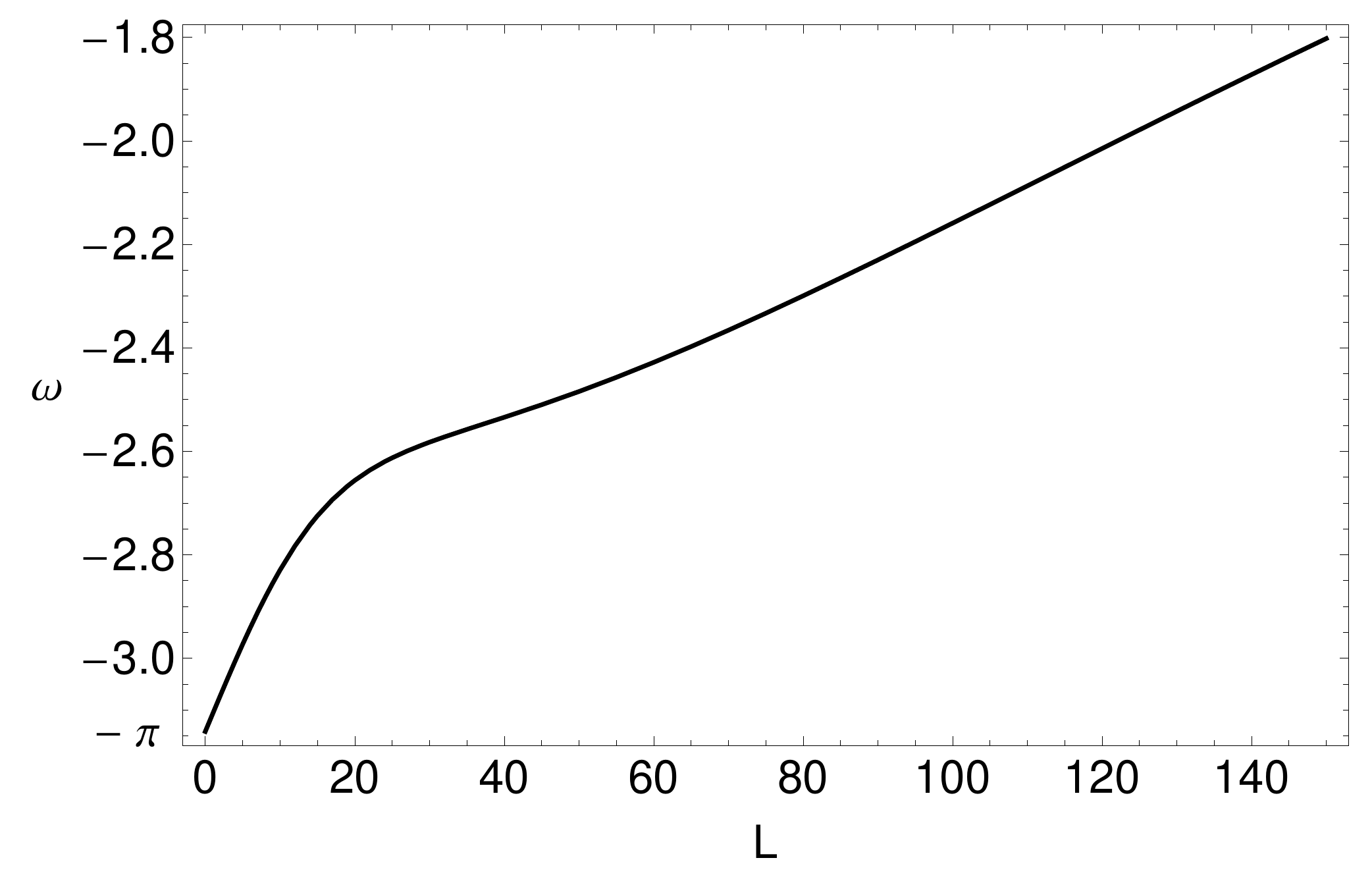}
\end{center}
\caption{The rotation angle $\omega$ as a function of the angular momentum $L$.}
\label{rot}
\end{figure}

For the case of $L = 0$ shown in part (a) of Fig. \ref{map1} any trajectory moves in a 2 dimensional plane in position space which contains the $z$-axis. In addition for $\alpha = 1$ the motion is independent of the angle $\phi$ of this plane. Therefore for $\alpha = 1$ and $L = 0$ $M_{\rm red}$ coincides with the maps $M_x$ and $M_y$.

Next we explain how we pile up the stack of reduced systems to represent the 4 dimensional Poincar\'{e} map. First, imagine that we construct the reduced map for all values of $L$ in the interval $[- L_{max}, L_{max}]$ and pile up this continuum of 2 dimensional maps to a 3 dimensional stack where the vertical coordinate (the stack parameter) is $L$. Second, we form the Cartesian product of this stack with a circle representing the cyclic angle $\phi$. Thereby we arrive at a 4 dimensional object. The natural coordinates of this object are $z$, $p_z$, $\phi$ and $L$, i.e. exactly the 4 coordinates of the domain of $M$, i.e. this 4 dimensional object can be identified with the domain of the 4 dimensional Poincar\'{e} map $M$. Finally, we have to understand how $M$ acts on this construction. In each copy of the $z$ - $p_z$ plane the map acts according to $M_{\rm red}$ for the respective value of $L$. $L$ itself is conserved. And each copy of the circle suffers a rigid rotation. Fig. \ref{rot} shows $\omega$ as function of $L$ for the trajectory moving in the equatorial plane $z = 0$. By choosing the equatorial plane we obtain exactly that twist curve which we need in a moment to understand the perturbed map $M_z$. It is only plotted up to a value of $L = 150$ since higher values are irrelevant for the real galaxy. However, the function continues up to $L_{\rm max}$ where the rotation angle reaches a value around -1.2. For negative values of $L$ the function should be continued antisymmetrically.

After having reached the global understanding of the 4 dimensional map for $\alpha = 1$ we have to study the perturbation scenario of this map, i.e. its behaviour under changes of $\alpha$, or even better we should use $\epsilon = \alpha - 1$ as perturbation parameter. The perturbation has a strong effect, if the $\phi$ dependence of the perturbation comes in resonance with the rotation angle $\omega$ of the unperturbed system. In our case the perturbation appears in the argument of the logarithm of the halo potential only and in this argument we set
\begin{equation}
x^2 + \alpha y^2 = R^2( 1 + \epsilon \sin^2(\phi))
\label{pert}
\end{equation}
In this sense the perturbation consists of a Fourier component of order 2 only and has most effect when the rotation angle $\omega$ has values around $\pi$. As Fig. \ref{rot} shows, this happens for $L$ values around 0. Therefore the perturbation will have strong effects around $L \approx 0$ also for small values of $\epsilon$. Higher order resonance effects are expected to be small for small values of $\epsilon$ and to become efficient for large values of $\epsilon$ only. The next important resonances are a 1:3 resonance for $L \approx 110$, a 1:4 resonance for $L \approx 185$ and a 3:8 resonance for $L \approx 70$. Corresponding resonances exist for negative values of $L$.

The perturbation scenario is presented in the most evident form by the restricted map $M_z$ on its domain with coordinates $\phi$ and $L$. These are the canonical coordinates of the $\phi$ degree of freedom and this degree of freedom is exactly the one which is decoupled from the rest of the system in the symmetric case. In the symmetric case $M_z$ is very trivial since $L$ is conserved and the domain is foliated into the invariant lines of constant $L$ (such lines are circles of the variable $\phi$, one for each value of $L$). Each of these circles is just rigidly shifted by the rotation angle $\omega(L)$. In this sense the map $M_z$ for $\alpha = 1$ is a pure twist map. Under perturbations any twist maps follows the universal perturbation scenario of the standard map \citep{C79}. For useful information on the interpretation of structures found in Poincar\'{e} maps see also chapter 4 in \citet{LL83}.

\begin{figure*}[!tH]
\centering
\resizebox{0.8\hsize}{!}{\includegraphics{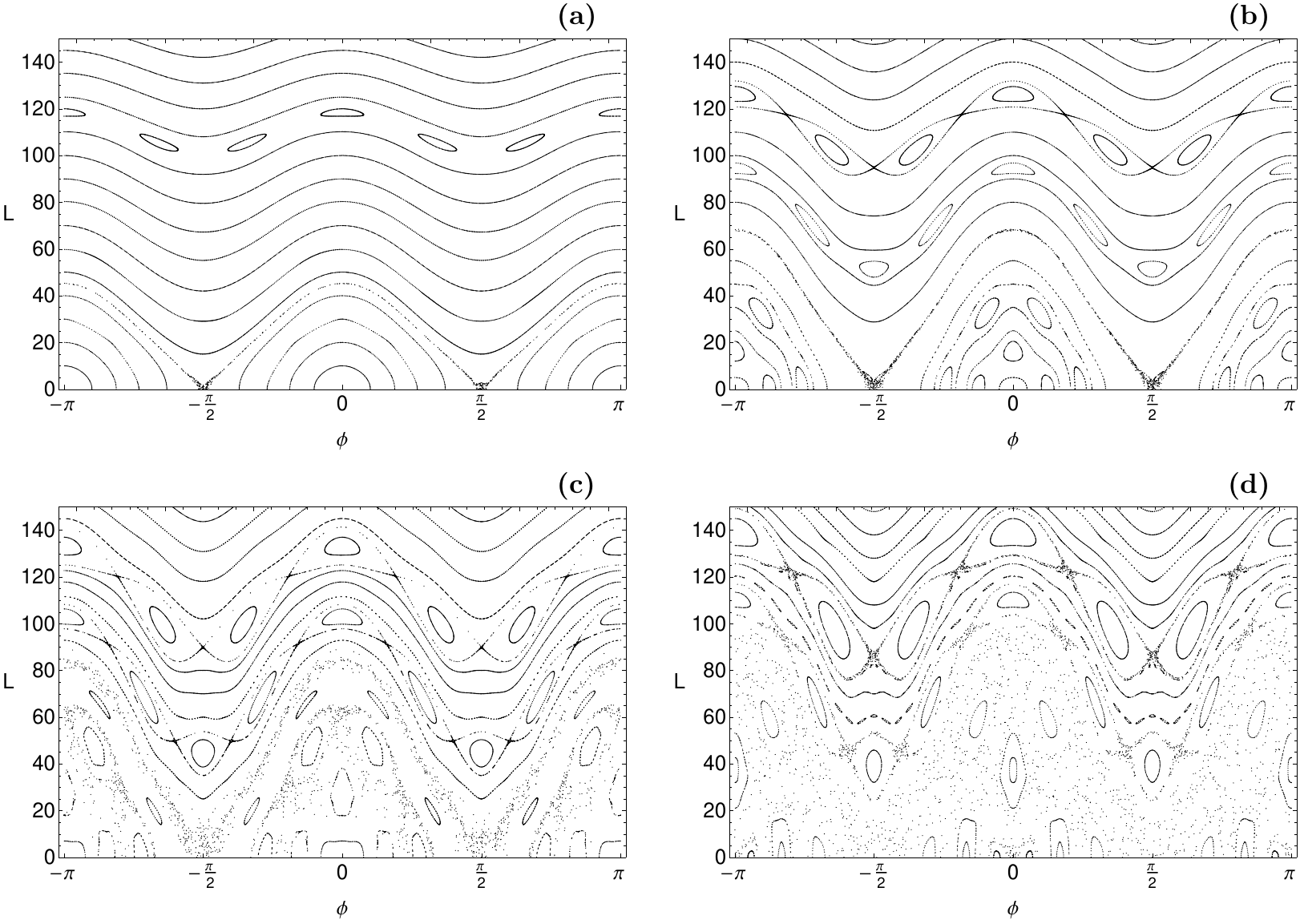}}
\caption{Examples of the perturbed map $M_z$ for various values of $\alpha$. (a) $\alpha = 1.25$; (b) $\alpha = 1.5$; (c) $\alpha = 1.75$; (d) $\alpha = 2$.}
\label{map2}
\end{figure*}

In Fig. \ref{map2} we show 4 examples for the perturbed map $M_z$. Only the part of the domain up to $L = 150$ is included in the plot since the model is only realistic up to this value. In principle the domain goes up to $L_{\rm max}$ and the natural boundary of the domain is the trajectory $\gamma_2$. The plots should be supplemented symmetrically for negative values of $L$.

For the moderate perturbation strength $\alpha = 1.25$ in part (a) of the figure the 1:2 resonance at $L = 0$ has already a very large effect, it causes two large secondary islands and a fine chaos strip which reaches up to $L$ values around 45. The centres of the secondary island are the two points $(0,0)$ and $(\pi,0)$ of an elliptic orbit of period two, and the centres of the chaos strips are the two points $(\pi/2,0)$ and $(-\pi/2,0)$ of a hyperbolic orbit of period two. The fine chaos strip still comes very close to a separatrix curve and it indicates the position of the initial segments of the stable and unstable manifolds of the unstable orbit of period two. The secondary structure of the 1:3 resonance at $L$ values around 110 consists of two distinct island chains of period 3. These two different island chains are transformed into each other by the discrete symmetry $\phi \to \phi + \pi$. The corresponding chaos strip is extremely fine and not displayed in the figure. All other secondary structures are still too small to become visible at this perturbation strength.

For $\alpha = 1.5$ in part (b) of the figure the secondary structure of the 1:2 resonance has grown to a width in $L$ of
approximately 70 and its chaos strip has become wider so that it is clearly distinct from a separatrix curve. Further, the fine chaos strip of the 1:3 resonance has become visible and the island chain of the 3:8 resonance has become visible. However, in total the map still is dominated by regular motion.

For the perturbation $\alpha = 1.75$ in part (c) of the figure the interior of the 1:2 resonance islands has become chaotic on a large fraction of the area. And we see the fine chaos strip belonging to the 3:8 resonance.

Finally in part (d) we present the case $\alpha = 2$. Here the 1:2 resonance island has already dissolved into a large chaotic sea with only a few surviving islands inside of the chaotic sea, they come from higher levels of the hierarchy. In addition the chaos strips of the 3:8 and of the 1:3 resonances have become so wide that we can distinguish them very clearly from separatrix curves. At least for $L$ values below 100 the map is now dominated by instability, whereas for large values of $L$ it is still mainly integrable. In total we see in Fig. \ref{map2} a perturbation scenario following the standard map as it is expected for any generic perturbed twist map. It should be emphasized that the dynamics in the surface $S_z$ is independent of $\beta$ and therefore Figs. \ref{rot} and \ref{map2} are equally valid for any other value of $\beta$.

\begin{figure*}[!tH]
\centering
\resizebox{0.8\hsize}{!}{\includegraphics{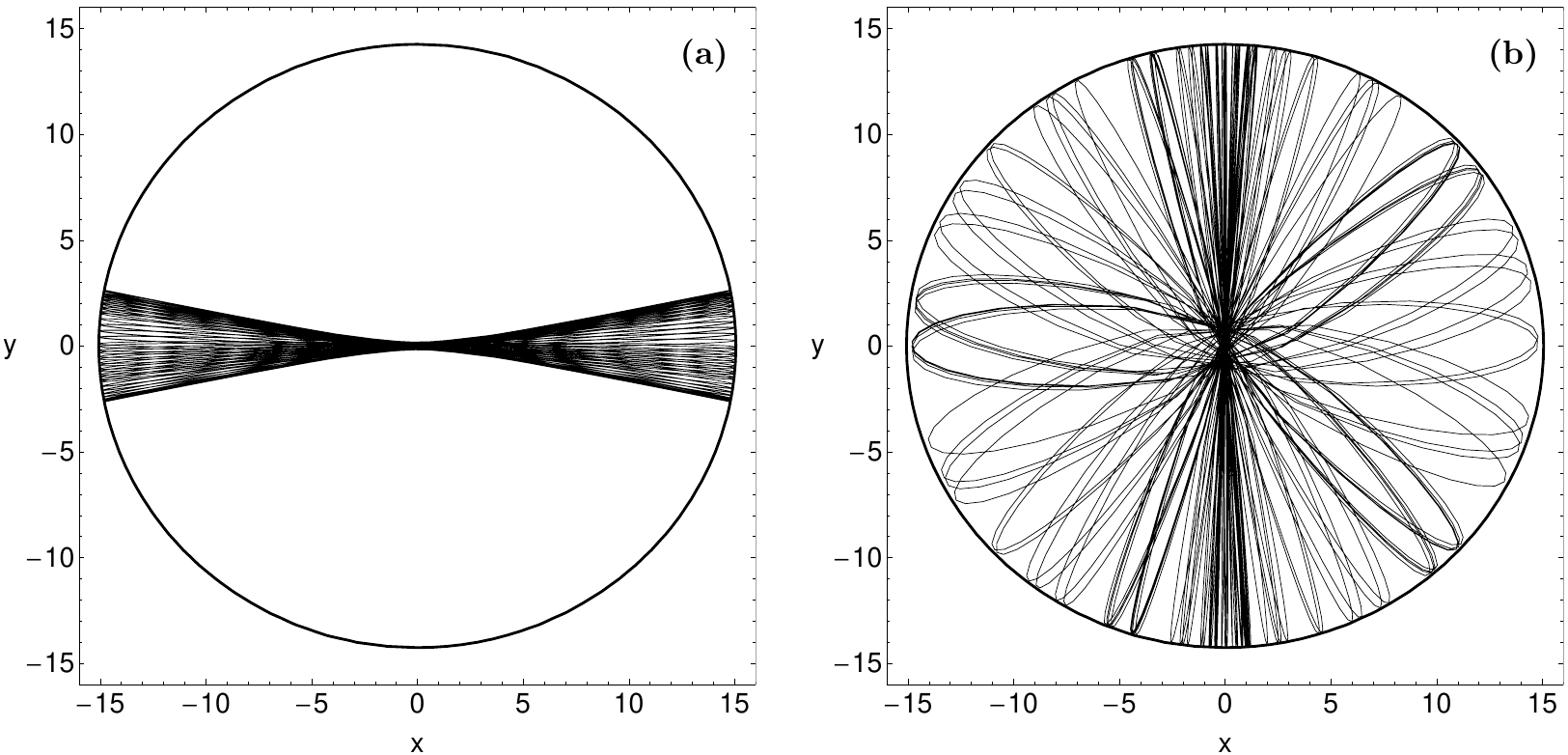}}
\caption{Two trajectories in the $(x,y)$ plane of the position space for parameter values $\alpha = 1.5$, $\beta = 0.1$ and $E = 160$. Part (a-left) belongs to the initial condition $\phi = 0.1$ and $L = 5$ which lies close to the elliptic point of period 2 in Fig. \ref{map2}b. Part (b-right) belongs to the initial condition $\phi = 1.57$ and $L = 2$ which lies close to the hyperbolic point of period 2 in Fig. \ref{map2}b.}
\label{orbs2}
\end{figure*}

It is seen in Fig. \ref{map2} that the most prominent centres for the perturbation scenario are the points of period 2 on the line $L = 0$. The elliptic points at $\phi = 0$, $L = 0$ and at $\phi = \pi$, $L = 0$ correspond to a trajectory in position space which oscillates along the $x$-axis. And the hyperbolic points at $\phi = \pm \pi/2$, $L = 0$ correspond to a trajectory in position space which oscillates along the $y$-axis. In Fig. \ref{orbs2} we show two trajectories in position space with initial conditions close to these two central trajectories. The outermost solid line is the Zero Velocity Curve at the $(x,y)$ plane defined as $\Phi(x,y,0) = E$. We choose the examples for the perturbation $\alpha = 1.5$ corresponding to part (b) of Fig. \ref{map2}. Also the nearby trajectories shown start with initial conditions $z = 0$ and $p_z = 0$, they lie in the invariant plane $S_z$ and thereby represent points in the Poincar\'{e} plot of Fig. \ref{map2}b. The trajectory plotted in Fig. \ref{orbs2}a has initial conditions $\phi = 0.1$, $L = 5$. Because it represents the corresponding point in Fig. \ref{map2}b it must have the additional initial condition $p_R = 0$ and the initial value of $R$ is given by the energy value $E = 160$ resulting in $R \approx 15.05$. Fig. \ref{orbs2}a is the resulting trajectory in the equatorial plane $S_z$ plotted in coordinates $x$ and $y$. It represents a small KAM curve in Fig. \ref{map2}b encircling the elliptic point of period 2. It is the smallest KAM curve around the elliptic point included in Fig. \ref{map2}b. The direction of oscillation of this trajectory fluctuates quasi-periodically around the values $\phi = 0$ or equivalently $\phi = \pi$. The width of these fluctuations in $\phi$ is given by the horizontal width of the corresponding KAM curve in Fig. \ref{map2}b. The trajectory plotted in Fig. \ref{orbs2}b has initial conditions $\phi = 1.57$, $L = 2$. Also it has the additional initial condition $p_R = 0$ and the initial value of $R$ is given by the energy value $E = 160$ resulting in $R \approx 14.25$. Fig. \ref{orbs2}b is the resulting trajectory in the equatorial plane with coordinates $x$ and $y$. It represents motion in the small chaos strip in Fig. \ref{map2}b containing the hyperbolic point of period 2. The direction of oscillation of this trajectory starts in $y$ direction, i.e. around the values $\phi = \pm \pi/2$ with the initial value 2 of $L$. Then the direction slowly rotates away from the $y$ direction passes through the $x$ direction quite rapidly and returns slowly to the $y$ direction before making the next rotation of the direction of oscillation. Because it is a chaotic trajectory the rotations of the direction of oscillation occurs irregularly and also the orientation of rotation changes in the long run again and again in an irregular fashion. When the direction of oscillation lies in the $x$ direction this corresponds to a point in the small chaos strip of Fig. \ref{map2}b at $\phi$ values close to 0 or $\pi$ and correspondingly the angular momentum of the trajectory lies around 70 at this moment. Accordingly the trajectory is more round at this moment.

\begin{figure*}[!tH]
\centering
\resizebox{0.8\hsize}{!}{\includegraphics{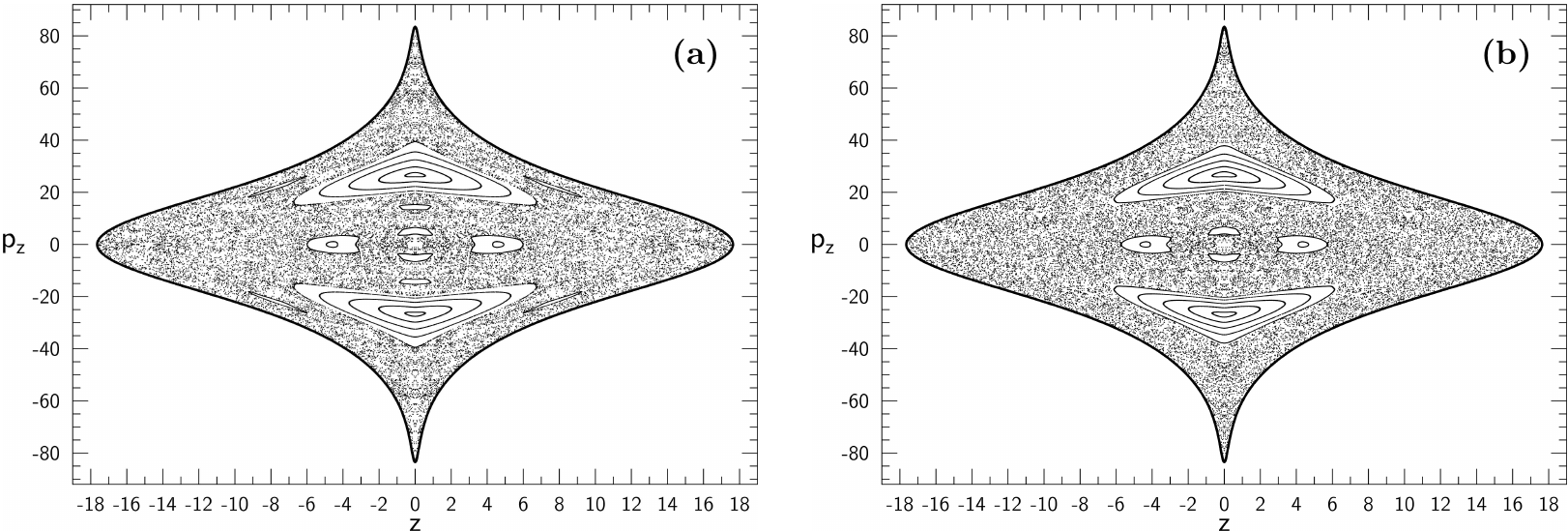}}
\caption{Plot of the Poincar\'{e} $M_x$ when $\beta = 0.1$ and (a-left): $\alpha = 1.5$ and (b-right): $\alpha = 2$.}
\label{map3}
\end{figure*}

To understand the 4 dimensional map $M$ for $\alpha \ne 1$ it is helpful to have in addition the restricted maps $M_x$ and $M_y$. First, in the surface $S_y$ the dynamics is independent of $\alpha$. Therefore the plot of the map $M_y$ always coincides with the plot of the map $M_{\rm red}$ for $L = 0$ as shown in Fig. \ref{map1}a. For $\alpha = 1$ also the map
$M_x$ coincides with the map of Fig. \ref{map1}a. However, for $\alpha \ne 1$ the map $M_x$ shows a weak dependence on $\alpha$. In Fig. \ref{map3} we show the cases of $M_x$ for $\alpha = 1.5$ and $\alpha = 2$ in parts (a) and (b) respectively. By comparing Figs. \ref{map1}a, \ref{map3}a and \ref{map3}b we see that there is a rather small dependence of $M_x$ on $\alpha$, but only in fine details. The energetic boundaries of Fig \ref{map3} are given by the trajectory $\gamma_1$. For this trajectory along the $z$-axis all energy is in the $z$ motion and the dynamics along this line is independent of the value of $\alpha$.

Finally we intent an approximate global understanding of the 4 dimensional map $M$ for general values of $\alpha$. The plane $S_x$ can also be characterized as plane given by $L = 0$ and $\phi = \pm \pi/2$ and the plane $S_y$ is at the same time the subset characterized by $L = 0$ and $\phi = 0$ or $\phi = \pi$. The 4 dimensional domain of $M$ can be considered as Cartesian product of the $\phi$ - $L$ space with the $z$ - $p_z$ space. Over each point of the $\phi$ - $L$ space we imagine sitting one copy of the $z$ - $p_z$ space. Each one of the two copies of the $z$ - $p_z$ space sitting over the points $\phi = \pm \pi/2$, $L = 0$ can be identified with $S_x$. These fibers are invariant and the restriction of $M$ to these fibers
coincides with $M_x$. Equally each one of the two copies of the $z$ - $p_z$ space sitting over the points $\phi = 0$, $L = 0$ and $\phi = \pi$, $L = 0$ can be identified with $S_y$. These fibers are invariant and the restriction of $M$ to these fibers coincides with $M_y$. Then in the neighbourhood of such invariant planes we can imagine as a first approximation for the complete map $M$ the product of the restricted map $M_z$ and $M_x$ or $M_z$ and $M_y$, respectively. For general points in the 4 dimensional domain this is a rather crude approximation but it helps to get some first imagination of the approximate
action of the full map $M$ in 4 dimensions.

\begin{figure*}[!tH]
\centering
\resizebox{0.8\hsize}{!}{\includegraphics{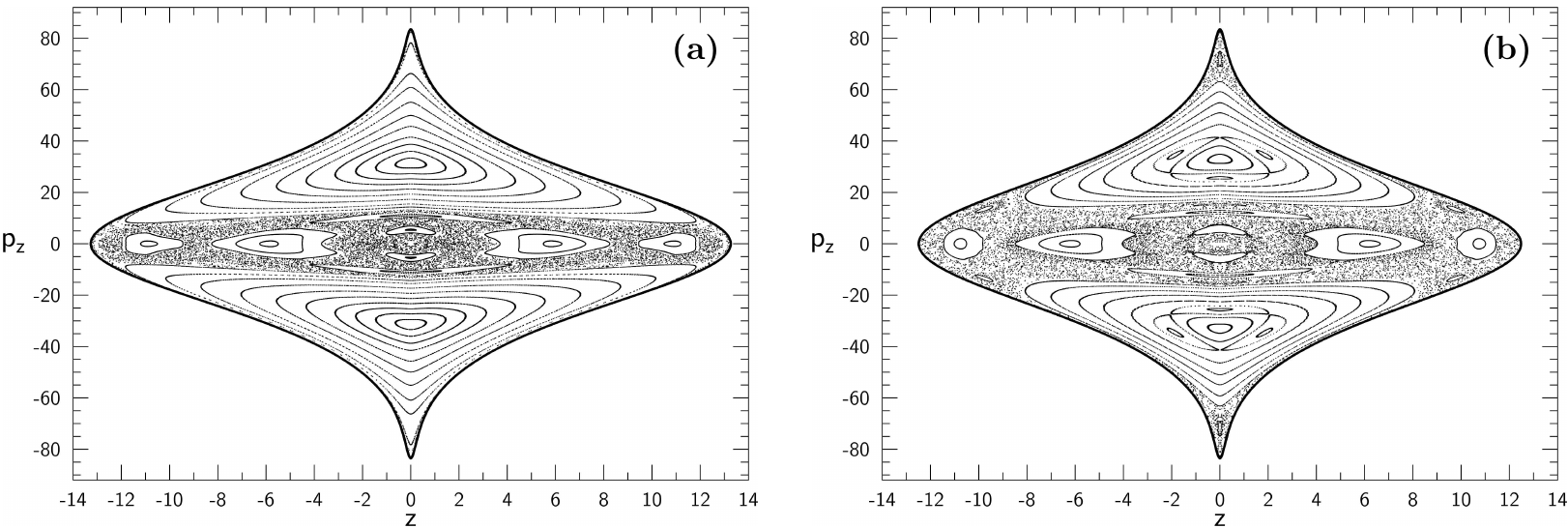}}
\caption{Plot of the Poincar\'{e} $M_x$ when $\alpha = 1$ and (a-left): $\beta = 1$ and (b-right): $\beta = 1.5$.}
\label{map4}
\end{figure*}

\begin{figure*}[!tH]
\centering
\resizebox{0.8\hsize}{!}{\includegraphics{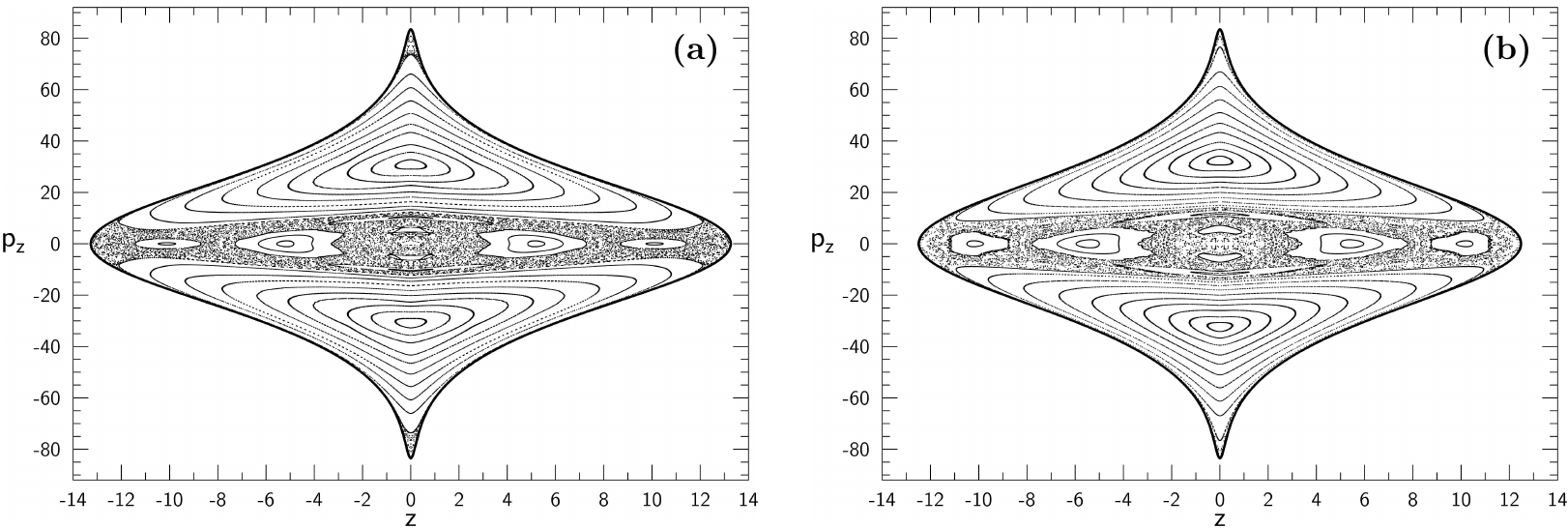}}
\caption{Plot of the Poincar\'{e} $M_x$ when $\alpha = 1.5$ and (a-left): $\beta = 1$ and (b-right): $\beta = 1.5$.}
\label{map5}
\end{figure*}
So far we have considered the case $\beta = 0.1$ only. In Figs. \ref{map4} and \ref{map5} we show the plots necessary to understand the cases $\beta = 1$ and $\beta = 1.5$. First we have to remember that the plots of Figs. \ref{rot} and \ref{map2} are independent of $\beta$. In Figs. \ref{map4} and \ref{map5} we show the plots of the map $M_x$ for the various combinations of $\alpha$ values of 1 and 1.5 and $\beta$ values of 1 and 1.5. Remember that the plot of $M_y$ always coincides with the corresponding $M_x$ plot for $\alpha = 1$. Thereby we have the necessary information to understand also the cases $\beta = 1$ and $\beta = 1.5$.

\begin{figure*}[!tH]
\centering
\resizebox{0.8\hsize}{!}{\includegraphics{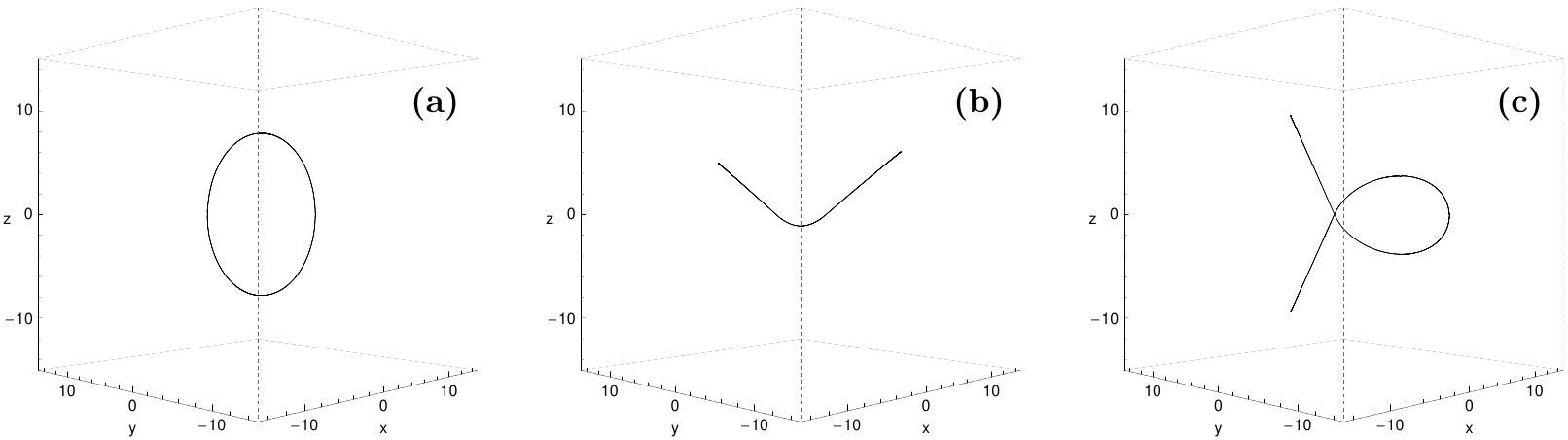}}
\caption{Orbit collection of the basic types of resonant periodic orbits in our galaxy model when $\alpha = \beta = 1.5$: (a) 0:1:1; (b) 0:2:1; (c) 0:3:2.}
\label{orbs3}
\end{figure*}

Before closing this section we would like to present in Fig. \ref{orbs3}(a-c) some examples of each of the basic types of regular resonant orbits. The orbit shown in Fig. \ref{orbs3}a is a 0:1:1 resonant periodic orbit with initial conditions $(z_0, p_{z0}) = (0, 32.54128711)$. A 0:2:1 resonant periodic orbit with initial conditions $(z_0, p_{z0}) = (5.44967215, 0)$ is given in Fig. \ref{orbs3}b, while the initial conditions of the 0:3:2 periodic orbit of Fig. \ref{orbs3}c are $(z_0, p_{z0}) = (10.18224703, 0)$. The $n:m:l$ notation we use for the regular orbits is according to \citet{CA98} and \citet{ZC13}, where the ratio of those integers corresponds to the ratio of the main frequencies of the orbit, where main frequency is the frequency of greatest amplitude in each coordinate. Main amplitudes, when having a rational ratio, define the resonances of an orbit. For the periodic orbits of Fig. \ref{orbs3}(a-c) we have $x_0 = 0, p_{xp} = 0$ because the trajectories lie in this invariant plane, while from the intersection condition we know that $p_{y0} = 0$ since for this particular situation $p_y$ coincides with $p_R$ and our intersection condition is $p_R = 0$. Finally the initial value of $y$, which coincides with $R$ in this case, has to be calculated by the total energy integral. The values of the parameters $\alpha$ and $\beta$ are as in Fig. \ref{map5}b.

\section{Discussion and conclusions}
\label{disc}

The purpose of the whole work is to obtain knowledge for which parameter values (here we think of parameters $\alpha$ and $\beta$ that control the geometry of the dark matter halo) and for which regions of phase space we have stability and instability and which types of trajectories play an important role.

An inspection of Figs. \ref{map1}, \ref{map2}, \ref{map3}, \ref{map4}, \ref{map5} demonstrates that the 1:2 resonance in the invariant plane $S_z$ (most clearly shown in Fig. \ref{map2}) plays the most prominent role. It depends strongly on $\alpha$ but is completely independent of $\beta$. This importance is easy to understand since the functional form of the perturbation enforces a 1:2 coupling with the $\phi$ degree of freedom, i.e. the degree of freedom which is decoupled from the rest of the dynamics in the partially integrable limit case. Because the resonance occurs around $L = 0$ it mainly affects trajectories passing close to the center of the galaxy. As can be seen from the functional form of the halo potential, the $y$ direction (or the angle values $\phi = \pm \pi/2$ as seen from Eq. (\ref{pert})) is the one most affected by the perturbation, i.e. by the breaking of the rotational symmetry. With increasing $\alpha$ the width of oscillations in $y$ direction shrinks. As seen in Fig. \ref{map2} for moderate values of $\alpha$ initial conditions close to $L = 0$ and $\phi = 0$ or $\phi = \pi$ lead to stable oscillations in $x$ direction. On the other hand, initial conditions close to $L = 0$ and $\phi = \pm \pi/2$ lead to unstable behaviour. The trajectory first oscillates in $y$ direction, then its direction of oscillation rotates and at the same time the angular momentum increases and becomes maximal when the trajectory oscillates in $x$ direction. Then the direction of $R$ oscillation keeps on rotating while the angular momentum decreases again until it returns to $L \approx 0$ when the direction of $R$ oscillation returns to the $y$ direction. This is the behaviour of trajectories in the chaos strip starting from the hyperbolic points $L = 0$, $\phi= \pm \pi/2$ in plot shown in Fig. \ref{map2}. For larger perturbation around $\alpha \approx 1.75$ the oscillation in $x$ direction becomes inverse hyperbolic, see the transition from Fig. \ref{map2}b over \ref{map2}c to \ref{map2}d. Then most motion with small values of $L$ becomes unstable.

On the other hand, the parts of Fig. \ref{map1} for larger values of $L$ and Fig. \ref{map2} show that motion with large values of $L$ always stays stable. This indicates that the approximately circular motion of the disc around the centre of the galaxy is always stable. For larger values of $\alpha$ the secondary stable structures of the 1:3 resonance and the 3:8 resonance become moderate size. This leads to stable bundles of trajectories in the neighbourhood of the corresponding
periodic orbits.

Influences of the parameter $\beta$ can only exist for trajectories which go far outside of the galactic plane $z = 0$. The most extreme case is the trajectory $\gamma_1$. Figs. \ref{map4} and \ref{map5} suggest the following: Remember that the intersection condition is $p_R = 0$. Then points in Figs. \ref{map4} and \ref{map5} with large values of $p_z$ are trajectories which pass through the maximal value of $R$ when at the same time they have a large momentum in $z$ direction. And these are trajectories having their major part of the energy in the $z$ motion. Figs. \ref{map4} and \ref{map5} indicate that such trajectories are regular for large values of $\beta$. The situation is different for small $\beta$ as is evident from Fig. \ref{map1}a. Here we have stable islands of moderate size only. And in particular, we see chaos close to the whole boundary. This indicates that the trajectory $\gamma_1$ has become unstable. This is understandable: A small $\beta$ allows very wide oscillations in $z$ direction, wider than in $x$ or $y$ directions. This leads to overfocusing of the trajectory $\gamma_1$ and therefore to inverse hyperbolic behaviour (e.g., \citet{H04b,KDP09,KMP08,LMJ09,PDK09}).

In order to have a more complete view on how energy influences the orbital structure of the dynamical system we constructed additional maps for lower and higher values of the standard $E = 160$ level. Our numerical calculations suggest that for low energies, that is the case of local motion around the nucleus, the trajectories do not go out into the dark matter halo and therefore the values of the halo parameters are irrelevant. Whereas for large values of the energy the trajectories go out far and are strongly influenced by geometrical details of the halo.

\section*{Acknowledgment}

This work has been supported by DGAPA under grant number IG-101113 and by CONACyT under grant number 79988. The authors would also like to thank the anonymous referee for the careful reading of the manuscript and for all the apt suggestions and comments which allowed us to improve significantly both the quality and the clarity of the paper.

\section*{Appendix}
\label{apex}

When constructing a Poincar\'{e} map it would be of particular interest to know how much area on the phase plane is covered by initial conditions that correspond to chaotic orbits. Here we are going to present a simple but yet very effective numerical method for determining the chaotic percentage in Poincar\'{e} maps. To begin with it is important to have a Poincar\'{e} map constructed in fine detail. The density of points must be high enough that the following two conditions are fulfilled: First, in chaotic seas almost all pixels must be occupied. Second, KAM lines must appear as continuous curves without interruptions. After obtaining the data points from the numerical integration routine we load them into the software program {\it Mathematica} \citep{W03} and we plot them thus creating the map.

\begin{figure}[!tH]
\begin{center}
\includegraphics[width=\hsize]{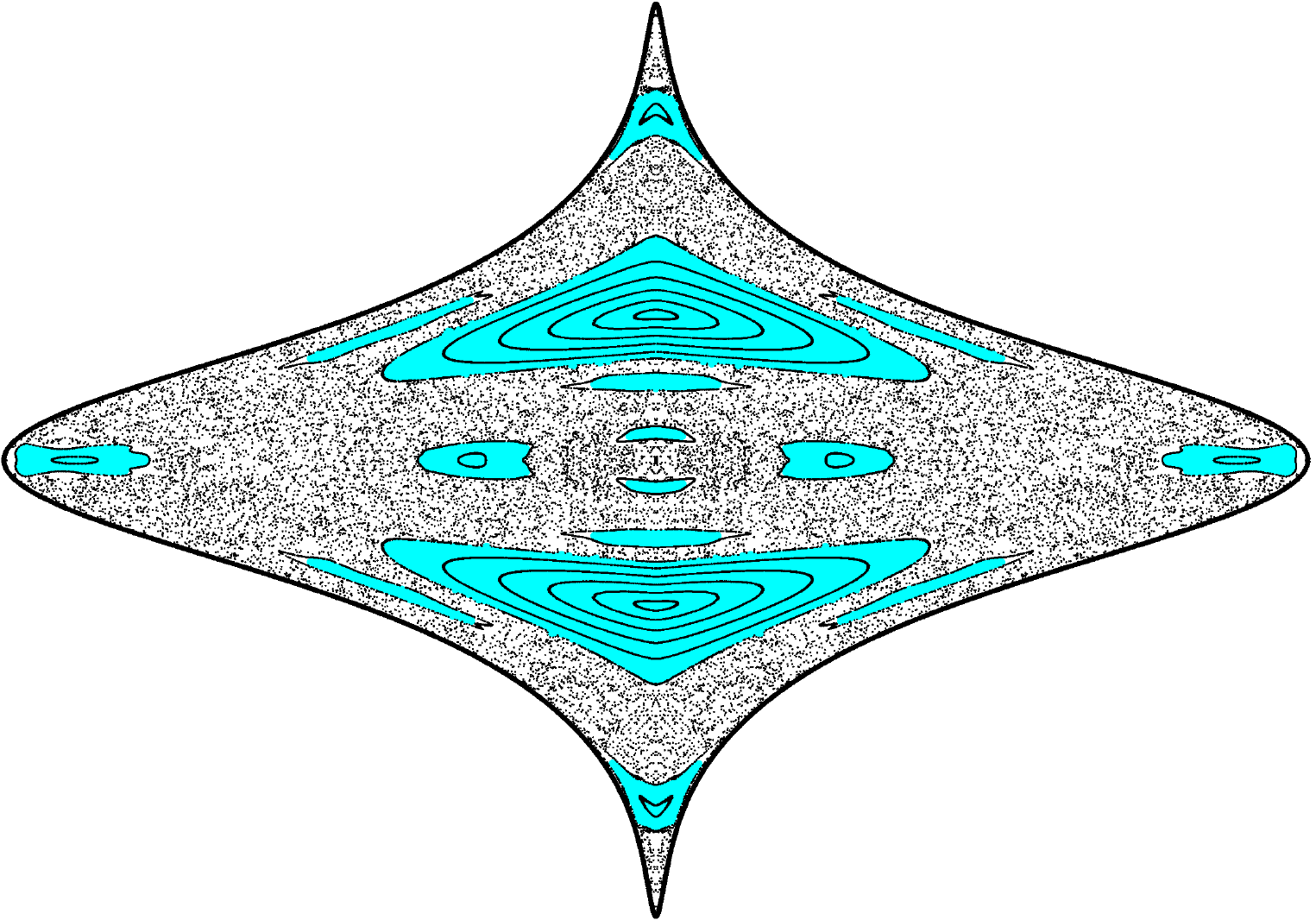}
\end{center}
\caption{The Poincar\'{e} map of Fig. \ref{map1}a combined with the stability islands (cyan) detected by the image-processing method.}
\label{img}
\end{figure}

This software program contains a plethora of image-processing tools. By using the appropriate functions we locate all the white holes in the map which correspond to stability islands of regular orbits. Then the map is divided into two parts: (i) the mask of the full available region on the phase plane and (ii) the mask of the regular regions. For both parts we measure how many pixels each mask covers. Therefore, if we assume that the mask of the full region occupies $N$ pixels and the mask of the regular regions occupies $n$ pixels then the chaotic area on the phase plane is $(N - n)/N$. In Fig. \ref{img} we present an illuminating example combining the detected stability islands with the full Poincar\'{e} map of Fig. \ref{map1}a in an attempt to to point out the great efficiency of the numerical method.

\end{document}